\documentstyle[times,balanced,epsf,pre,aps,amsfonts]{revtex}

\begin{document}

\draft

\title{
Emergent Global Oscillations in Heterogeneous Excitable Media; \\
The Example of Pancreatic $\beta$ Cells
}

\author{
Julyan H. E. Cartwright\cite{jemail}
} 

\address{
Laboratorio de Estudios Cristalogr\'aficos, IACT (CSIC-UGR), 
E-18071 Granada, Spain.
}

\date{Phys.\ Rev.\ E, {\bf 62}, 1149--1154, 2000
}

\maketitle

\begin{abstract}
Using the standard van der Pol--FitzHugh--Nagumo excitable medium model I 
demonstrate a novel generic mechanism, diversity, that provokes the emergence 
of global oscillations from individually quiescent elements in heterogeneous 
excitable media. This mechanism may be operating in the mammalian pancreas,
where excitable $\beta$ cells, quiescent when isolated, are found to oscillate
when coupled despite the absence of a pacemaker region.
\end{abstract}


\begin{twocolumns}

\section{Excitable Media}

An excitable element is defined by its response to a perturbation: whereas a
small disturbance causes merely an equally small response, a perturbation above
a certain threshold in amplitude excites a quiescent element that then decays
back to quiescence during a refractory period in which it is unresponsive to
further excitation. Such elements, when coupled to their neighbors into an
assembly, become an excitable medium \cite{meron}. These have attracted an
enormous amount of interest from different areas of science: the
Belousov--Zhabotinsky reaction \cite{zaikin}, plankton populations 
\cite{truscott}, and the heart \cite{glass,panfilov} are just a few well-known
examples. Oscillations in excitable media such as the heart are produced by
forcing the medium from a pacemaker region; in other cases all individual
elements of the medium oscillate when isolated: they are all pacemakers, and
the medium is then oscillatory rather than excitable.
However, this need not necessarily be so. In this paper I present a novel
generic mechanism for the production of global oscillations; the
introduction of diversity amongst the elements leads to the destabilization of
the quiescent state of an excitable medium and to the emergence of global
oscillations even when each individual element of the medium is quiescent in
isolation. I argue that an instance of such behavior may be found in a
physiological example of an excitable medium without a pacemaker: the
pancreatic $\beta$ cells of mammals.

In the mammalian pancreas are encountered structures where insulin is produced.
These, the islets of Langerhans, consist of several thousand spherical cells
clustered together and electrically connected
via resistive gap junctions \cite{meissner}. The vast majority of these
cells, those that produce insulin in response to the level of glucose in the
blood, are of a type known as $\beta$ cells. In
many mammals, including humans, the electrical potentials of $\beta$ cells in
an islet are found by experiment to cycle synchronously in slow oscillations
termed bursts \cite{dean}. On the other hand, an individual $\beta$
cell when removed from the islet does not oscillate in this way \cite{perez},
but rather is excitable. An islet of Langerhans, then, like the heart, is a
physiological excitable medium. But in the pancreas, unlike in the heart,
oscillations of the excitable medium are not driven by pacemaking cells. The
challenge is to understand the origin of these oscillations. Various detailed
biophysical models of oscillations in networks of pancreatic $\beta$ cells have
been proposed, in which the importance of the heterogeneity of individual cells
\cite{smolen} and the emergence of oscillatory behavior upon coupling
nonoscillatory cells \cite{sherman,andreu} have been highlighted. Here I take
a different approach in which I bring together these two ideas while
simplifying the physics as far as possible to produce a minimal qualitative
model for the phenomenon. I demonstrate that given diverse excitable elements, 
coupling these into a heterogeneous excitable medium can lead to the emergence 
of oscillations: that diversity is a generic mechanism for the emergence of 
global collective behavior not just in $\beta$ cells, but in any heterogeneous
excitable medium. 

\section{van der Pol--FitzHugh--Nagumo model}

\begin{figure}
\begin{center}
\def\epsfsize#1#2{0.48\textwidth}
\leavevmode
\epsffile{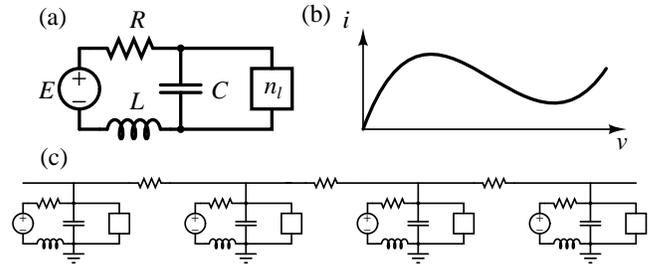}
\end{center}
\caption{\label{circuit}
An electronic excitable medium.
(a) An element of the medium: the circuit and (b) the $i$--$v$ characteristic of
the nonlinear element $n_l$: the cubic function $i\propto v^3/3-v$.  
(c) Resistive (diffusive) spatial coupling, illustrated for simplicity here 
in one dimension.
}
\end{figure}

To illustrate the mechanism, and having in mind its application to pancreatic
$\beta$ cells, I take an electronic circuit model as a caricature of a
physiological excitable medium and as representative of excitable media in
general. Each element of the medium is a circuit shown in
Fig.~\ref{circuit}(a). In physiological terms, the capacitor $C$ represents a
cellular membrane charged by $E$, characterizing ion pumps, and drained by a
nonlinear resistance $n_l$ across $C$. This nonlinear element, a device with a
range of negative resistance, could be a tunnel diode, for example, and should
have the cubic $i$--$v$ characteristic of  Fig.~\ref{circuit}(b). The
inductance $L$ models the finite switching time of the ion channels in the
membrane. The circuit is then mathematically described by the van der
Pol--FitzHugh--Nagumo equations 
\cite{vdp,fitz1,fitz2,nagumo}
\begin{mathletters}
\begin{eqnarray}
\dot\psi&=&\gamma(\eta-\psi^3/3+\psi), 
\label{eq1} \\
\dot\eta&=&-\gamma^{-1}(\psi+\nu+\beta\eta)
\label{eq2}
.\end{eqnarray}\label{eq}
\end{mathletters}
Variables $\psi$ and $\eta$ are respectively proportional to the potential
difference across the nonlinear device $n_l$ and the current through the supply;
parameter $\nu$ is proportional to the potential $E$, $\beta$ to the resistance
$R$, and $\gamma$ to the quotient $L/C$.
The equations have an equilibrium point
that is stable for $|\nu|>\Xi$ and unstable for $|\nu|<\Xi$, where 
\begin{equation}
\Xi=\sqrt{\gamma^2-\beta}\,\frac{3\gamma^2-2\gamma^2\beta-\beta^2}{3\gamma^3}
,\label{threshold}\end{equation}
so a circuit element is oscillatory for $|\nu|<\Xi$ and excitable for 
$|\nu|>\Xi$ in the vicinity of $|\nu|=\Xi$ \cite{winfree}.
In physiological terms, oscillatory behavior corresponds to bursting
($|\nu|<\Xi$), and excitability to silent and continuously 
active cells; one case being $\nu<-\Xi$ and the other $\nu>\Xi$.

Each of these elements is coupled to its nearest neighbors in one, two, or
three dimensions to become an excitable medium. The coupling in biological and
chemical excitable media is diffusive, though elastic excitable media that
arise in electronics and rheology have also recently been considered
\cite{quakeletter,quakebc}. Either or both of Eqs.\ (\ref{eq}) may host a
coupling term, depending on the medium being modeled. Here, to imitate a
cellular excitable medium, the circuits are coupled resistively as shown in
Fig.~\ref{circuit}(c). This leads to a diffusive term in Eq.\ (\ref{eq1})
\begin{equation}
\dot\psi=\gamma(\eta-\psi^3/3+\psi+\kappa\sum_{i=1}^n(\psi_i-\psi))
,\label{dif}\end{equation} 
where the $i$s represent the $n$ neighboring elements, and $\kappa$ is the
coupling strength or diffusion coefficient. In the continuum limit, the 
coupling term $\kappa\sum_{i=1}^n(\psi_i-\psi)$ becomes the Laplacian 
$\kappa\nabla^2\psi$; this is the
classical van der Pol--FitzHugh--Nagumo model of an excitable medium,
extensively analyzed in the excitable spiral-wave regime \cite{winfree}.

\section{Heterogeneous Excitable Media}

Consider now what happens if we introduce a diversity of parameter values for 
the different elements of the medium. The position of the stable equilibrium
point for Eqs.\ (\ref{eq}) depends on the parameters $\nu$ and $\beta$, so if
we introduce a spread of parameter values across the medium, we change the
equilibria of individual elements and the coupling between them
$\kappa\sum_i(\psi_i-\psi)$ will no longer be zero in the quiescent state. In
such a heterogeneous medium, the dynamics of each element may be analyzed to a
good approximation by treating the influence of the rest of the medium as a 
type of external signal $\varepsilon$. This is equivalent to setting
$\varepsilon=\kappa\sum_{i=1}^n(\psi_i-\psi)$ whence the extra term
$\varepsilon$ may be removed from Eq.\ (\ref{dif}) by renormalizing $\nu$, and
the element is modeled by Eqs.\ (\ref{eq}) with an effective $\nu$, 
$\nu'=\nu-\beta\varepsilon$; the medium is now oscillatory for 
$|\nu-\beta\varepsilon|<\Xi$. Hence a small $\varepsilon$ can push an excitable
element with $|\nu|>\Xi$ near to the excitable--oscillatory threshold over into
the oscillatory regime. This diversity mechanism is not specific to the van der
Pol--FitzHugh--Nagumo model, but rather is generic; it can be applied to any
heterogeneous excitable medium in which the position of equilibrium is
parameter dependent and there is an oscillatory regime reachable in the
extended parameter space formed by an uncoupled element's parameters plus 
the extra term $\varepsilon$.

\begin{figure}
\begin{center}
\def\epsfsize#1#2{0.48\textwidth}
\leavevmode
\epsffile{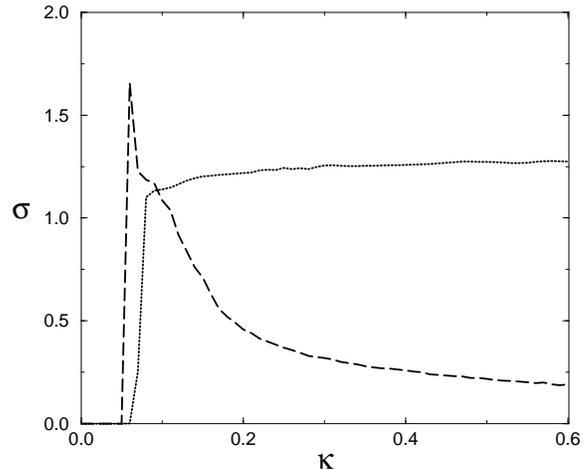}
\end{center}
\caption{\label{diversity}
Dotted line: Temporal standard deviation of $\psi$, $\sigma_t$  (Eq.\
(\ref{sigma_t})), shows the emergence of oscillatory behavior from a
heterogeneous excitable medium with increasing coupling $\kappa$. Dashed line:
Standard deviation of times $t_{max}$ of maxima of $\psi$, $\sigma_s$  (Eq.\
(\ref{sigma_s})), demonstrates increasing synchronization of the oscillations
with increasing coupling $\kappa$. The numerical results are for a van der
Pol--FitzHugh--Nagumo medium with  $\beta=0.5$, and $\gamma=2$, with $4\times
4\times 4$ elements randomly assigned $\nu=0.76$ or $\nu=-0.76$, both of which
are individually parameter values for which the medium is excitable rather than
oscillatory, and thus quiescent unless excited.
}
\end{figure}

Let us examine the emergence of oscillations through diversity in a simple
example using the van der Pol--FitzHugh--Nagumo model. In Fig.~\ref{diversity}
is plotted (dotted line) against coupling $\kappa$ the temporal standard 
deviation of $\psi_j$ for a sample element $j$, 
\begin{equation}
\sigma_t=\sqrt{\frac{1}{t_f-t_i}\sum_{t=t_i}^{t_f}\psi_j^2(t)-\bar\psi_j^2}
,\label{sigma_t}\end{equation}
for a heterogeneous van der Pol--FitzHugh--Nagumo medium consisting of elements
randomly assigned the parameter values $\nu=0.76$ or $\nu=-0.76$, together with
the other parameters $\beta=0.5$, $\gamma=2$. Both of these parameter sets on
their own produce excitable rather than oscillatory elements, which are
quiescent without excitation, so at $\kappa=0$, $\sigma_t$ is zero. But in 
Fig.~\ref{diversity} we see that as the coupling $\kappa$ between elements
passes a threshold, $\sigma_t$, which is measuring the temporal activity of the
medium, increases from zero, meaning that the medium has spontaneously begun to
oscillate; diversity has provoked the emergence of global oscillations from
individually quiescent elements. In this example, each element is connected in
a cubic lattice with six neighbors. Elements throughout the lattice are
randomly assigned one of two parameter values: $\nu=0.76$ or $\nu=-0.76$. On
average, any element will find  that half its neighbors share the same
parameter value, and half have the other value. At quiescence, the coupling
between those with the same parameter values is zero, while between those with
different parameter values it is
$\varepsilon=\kappa\sum_{i=1}^m(\psi_i-\psi)=3\kappa\Delta\psi$, since $m=3$ is
the average number of neighbors with different parameter values. In the van
der Pol--FitzHugh--Nagumo model, the change in the equilibrium value of $\psi$,
$\Delta\psi\approx-\Delta\nu$. For our example, $\nu=\pm 0.76$, which makes
$\Delta\nu=2\nu$. In this case, then, the new effective $\nu$ is
\begin{equation}
|\nu'|=|\nu-\beta\varepsilon|=|\nu(1-6\beta\kappa)|
.\label{nuprime}\end{equation}
We can see in  Fig.~\ref{diversity} that the medium begins to oscillate when
$\kappa\approx 0.06$. For the numerical values $\beta=0.5$, $\gamma=2$,
$|\nu|=0.76$, used in Fig.~\ref{diversity}, $|\nu'|=0.62$ when $\kappa=0.06$,
which corresponds closely to the threshold for oscillation in a homogeneous van
der Pol--FitzHugh--Nagumo medium given by Eq.\ (\ref{threshold}) for these 
parameter values: $|\nu|=\Xi=31/96\,\sqrt{7/2}=0.60412\ldots$. 

To simplify the above analysis as far as possible, I have considered a
heterogeneous medium with just two states. The diversity mechanism works in the
same way in a medium with a continuum of states, in which a proportion of
elements --- those whose uncoupled state falls within the parameter range for
autonomous oscillation --- may be intrinsic oscillators. However, such
intrinsic oscillators do not act like the pacemaker region of a driven medium.
Scattered randomly throughout a quiescent medium, they are neither sufficient
nor necessary for global oscillations.  Whether or not such intrinsic
oscillators are found in a heterogeneous medium depends on whether an
oscillatory regime exists within the parameter range of heterogeneity for an 
element when the system coupling term $\varepsilon$ is zero. In the van der
Pol--FitzHugh--Nagumo model this corresponds to whether the inequality
$|\nu|<\Xi$ is satisfied without coupling. For $\beta$ cells, the majority of
experiments have found no evidence for intrinsic oscillations, and that
isolated cells are exclusively excitable: either silent, or continuously active
\cite{rorsman}. 

\section{Synchronous and Asynchronous Oscillations}

The diversity mechanism applies very naturally to physiological excitable
media, since homogeneous cells are a mathematical fiction. The underlying
dynamics is such that there is a threshold for a quiescent heterogeneous medium
to achieve criticality and cross the
excitable--oscillatory boundary. This demands sufficient connectivity between
neighboring elements in terms of the number of connections (dimensionality of
the system) and the coupling strength (diffusion coefficient), together with
sufficient heterogeneity between elements. While the threshold-crossing
mechanism of diversity is generic, what happens on the other side of the
threshold once the medium has become oscillatory is not. In some oscillatory
media, synchronous global oscillations are stable, whereas in others these are
unstable to spatial perturbations leading to the formation of propagating
fronts or other spatial phenomena \cite{kramer}. The standard deviation across
the medium of times $t_{max}(k)$ of maxima of $\psi_k$ for elements $k$, 
\begin{equation}
\sigma_s=\sqrt{\frac{1}{N}\sum_{k=1}^{N}t_{max}^2(k)-\bar t_{max}^2}
,\label{sigma_s}\end{equation}
measures the spatial activity of the medium: the smaller this quantity, the
greater the synchronization throughout the medium. In our example, $\sigma_s$
plotted (dashed line) against coupling $\kappa$ in Fig.~\ref{diversity}
indicates that at the minimum coupling necessary for oscillatory behavior, the
heterogeneous medium oscillates with little synchronization. As the coupling is
increased, however, the synchronization rapidly improves, shown by the
steep decay in the standard deviation of the maxima at larger $\kappa$.

\begin{figure}
\begin{center}
\def\epsfsize#1#2{0.48\textwidth}
\leavevmode
\epsffile{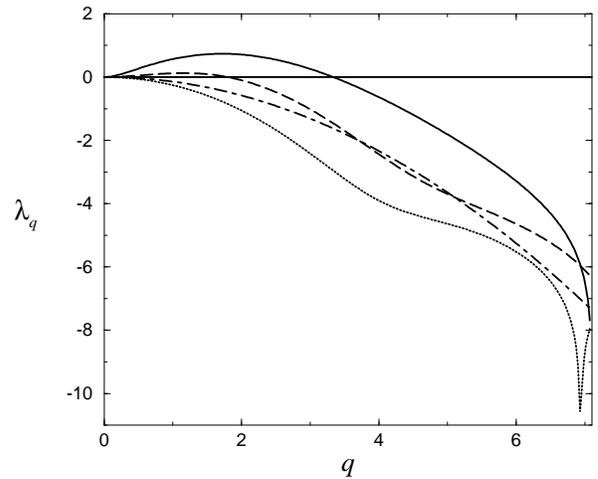}
\end{center}
\caption{\label{floquet}
Plots of dispersion relations show destabilization of synchronous global 
oscillations in the van der Pol--FitzHugh--Nagumo model for a range of $\nu$
in the oscillatory regime. Here $\beta=0.5$, $\gamma=2$, $\kappa=0.09$: shown
are curves for $\nu=0.592$ (dotted), $\nu=0.596$ (dashed), $\nu=0.600$
(solid), and $\nu=0.604$ (dot-dashed). The maximal Floquet exponent
$\lambda_q$ gives the growth rate per period of a perturbation of wavenumber
$q$ \protect\cite{quakeletter}. For $\nu=0.596$ and $\nu=0.600$, the
dispersion relation $\lambda_q$ as a function of $q$ is positive for a range
of $q$, meaning that there is a set of wavelengths for which perturbations
grow exponentially. Outside this range of $\nu$ values, the dispersion
relation is never positive, implying linear stability of synchronous global
oscillations under perturbations of all wavelengths. 
}
\end{figure}

If the synchronized state is an attractor for the homogeneous medium, then we
might expect the behavior in the heterogeneous case to reflect this. 
Let us consider for a moment the related area of synchronization, phase- and
frequency-locking, or entrainment of coupled oscillators, a vast field of study
initiated by Huygens with his observations of synchronization of two pendulum
clocks coupled by a common mounting \cite{huygens,huygens1,huygens2}.
Winfree \cite{winfree2} showed that synchronization emerges in a population of
heterogeneous oscillators as coupling exceeds a critical threshold, in a manner
reminiscent of a thermodynamic phase transition. Following this work Kuramoto 
\cite{kuramoto,sakaguchi1,sakaguchi2} developed his theoretical model whose
tractability helped to advance significantly the study of coupled heterogeneous 
oscillators. This undergoes two transitions as the spread of natural frequencies
is reduced, or the coupling is increased: first comes the onset of partial 
synchronization, which is followed by complete synchronization 
even with some residual heterogeneity among the elements. More recently a 
physically realizable version of the Kuramoto model has been proposed: an 
array of heterogeneous Josephson junctions \cite{wiesenfeld}. Strong coupling 
between oscillators allows modification of the amplitude as well
as the phase of an oscillator, which can give rise to novel phenomena such as
amplitude death \cite{bar-eli,ermentrout,mirollo} (here however we are 
in the weak coupling regime). Finally, if synchronous global
oscillations are unstable in the homogeneous case, the addition of a certain
amount of heterogeneity can even lead to an increase in synchronization, as
studies with locally coupled limit-cycle oscillators have found
\cite{braiman1,braiman2}. 

\begin{figure}
\begin{center}
\def\epsfsize#1#2{0.48\textwidth}
\leavevmode
\epsffile{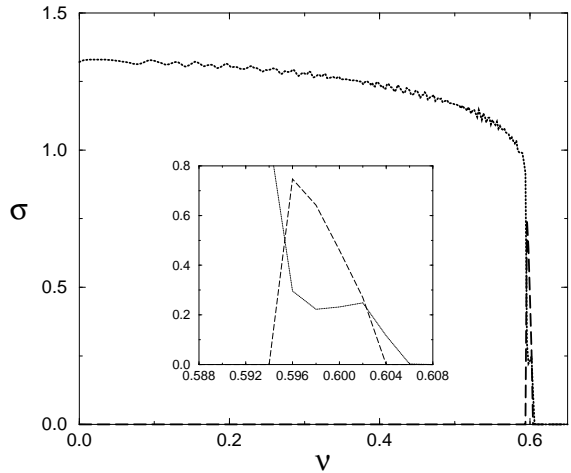}
\end{center}
\caption{\label{leakiness}
Dotted line: Temporal standard deviation of $\psi_j$, $\sigma_t$ 
(Eq.\ (\ref{sigma_t})), illustrates the transition to oscillatory behavior
in a homogeneous excitable medium with decreasing $\nu$.
Dashed line: Standard deviation of times $t_{max}$ of maxima of $\psi$, 
$\sigma_s$ (Eq.\ (\ref{sigma_s})), shows the destabilization of synchronous
global oscillations in a range of  $\nu$ just above the excitable--oscillatory
threshold. The inset highlights the narrow range $0.588<\nu<0.608$ in which the
destabilization occurs. Numerical results are for a van der
Pol--FitzHugh--Nagumo medium with  $4\times 4\times 4$ elements with
$\beta=0.5$, $\gamma=2$, and $\kappa=0.09$. From  Eq.\ (\ref{threshold}), the
medium is oscillatory for $|\nu|<\Xi=0.060412\dots$.
}
\end{figure}       

We can obtain a theoretical understanding of synchronization in a 
homogeneous oscillatory medium by calculating analytically the Floquet
exponents that indicate the linear stability of limit cycles against spatial
perturbations. This I have done for the van der Pol--FitzHugh--Nagumo model in
Fig.~\ref{floquet}, where I show a family of dispersion relations for different
values of $\nu$. A positive value for the maximal Floquet exponent illustrates 
the destabilization of synchronous global oscillations of the medium for a 
range of $\nu$ for a particular value of $\beta$. In the case shown, for which,
as in Fig.~\ref{diversity}, $\beta=0.5$, $\gamma=2$, the dispersion relation
$\lambda_q$ is positive for a range of wavenumbers $0<q<q_c$ for $\nu$ values
close to the excitable--oscillatory threshold of the medium. For these $\nu$
values then synchronous global oscillations are unstable to perturbations of
wavelengths $1/q_c<\lambda<\infty$ when these wavelengths fit within the system
size. For all other $\nu$ values in the oscillatory regime global synchronous 
oscillations of the homogeneous medium are linearly stable. This is
demonstrated numerically in Fig.~\ref{leakiness} in which are plotted against
$\nu$ the measures of activity and spatial synchronization $\sigma_t$ and
$\sigma_s$ for a homogeneous van der Pol--FitzHugh--Nagumo medium with 
$\beta=0.5$, $\gamma=2$, and $\kappa=0.09$. The dotted line $\sigma_t$ shows
that the medium is oscillatory up to $\nu=0.604$, then excitable, as Eq.\
(\ref{threshold}) demands. The dashed line $\sigma_s$ is positive near
$\nu=0.6$, showing that global oscillations are unstable for a range of $\nu$
values; outside this range the standard deviation $\sigma_s$ is zero, as the
medium oscillates synchronously. The inset highlights the range of $\nu$ in
which destabilization occurs.

In a homogeneous van der Pol--FitzHugh--Nagumo excitable medium we have seen
that there are parameter ranges in which synchronous global oscillations are
stable, and others in which they are not. The behavior of the homogeneous case
helps us now to understand the heterogeneous medium. Figure~\ref{diversity}
displays poor synchronization of the medium immediately following the emergence
of oscillations, which then improves with increasing $\kappa$: $\sigma_s$
peaks, then declines rapidly before leveling off. This can be seen as a
combination of the intrinsic dynamics of the van der Pol--FitzHugh--Nagumo
medium together with the  synchronizing effect of coupling. Upon the emergence
of oscillations in the heterogeneous medium, the effective $\nu$, $\nu'$ 
(Eq.\ (\ref{nuprime})), is within the range of values near to the
excitable--oscillatory transition for which synchronous oscillations are
unstable. As the coupling $\kappa$ increases, $\nu'$ drops below this range and
the synchronization immediately improves. The further slower decrease in
$\sigma_s$ is due to the additional synchronizing effect of coupling.

What might be the relevance of the synchronization of global oscillations for 
the $\beta$ cells of the pancreas? The pancreas may be contrasted with another
physiological excitable medium: the heart. While in the heart synchronous
oscillations are vital to the survival of the organism --- the unsynchronized
state of the fibrillating heart is fatal if not immediately resynchronized with
an electric shock --- it is not obvious why this should be so for $\beta$
cells. Although in humans and in mice the oscillations are synchronous, in
other mammals there is less evidence for this. While physiologists may yet 
provide a biological rationale for synchronization, it may be that it is not a
physiological necessity but simply a byproduct of the emergence of
oscillations: the pancreatic $\beta$ cells of some species may be operating in
a parameter range in which synchronous global oscillations are stable, while in
others, less investigated up to now, spatial patterns may be found in the
oscillations.

\section{Conclusions}

I have argued that diversity acts at a fundamental level in the dynamics of
heterogeneous excitable media to produce global oscillations. Another related
theme of study has subjected excitable systems to various types of forcing. 
A periodically forced van der Pol--FitzHugh--Nagumo element shows behavior
similar to a driven oscillator: phase locking, quasiperiodicity, period
doubling, and chaos \cite{fgpv}. A quiescent excitable element may be excited
by driving with a combination of a periodic subthreshold signal plus noise 
\cite{wiesenfeld2} or with an aperiodic subthreshold signal plus noise 
\cite{collins}, phenomena which have been termed stochastic resonance,
or with noise alone \cite{pikovsky}, when the phenomenon has been termed
coherence resonance. Here, we have seen that even without any external forcing, 
either periodic or stochastic, a heterogeneous excitable medium can become
self-excited to produce global oscillations.

To what extent is this general analysis applicable to pancreatic $\beta$ cells?
$\beta$ cells are diffusively coupled excitable elements, although more
complicated than van der Pol--FitzHugh--Nagumo elements, with more internal
variables and parameters. Their heterogeneity has been made manifest in studies
demonstrating differing rates among $\beta$ cells of insulin synthesis and
secretion \cite{salomon,stefan,schuit,hiriart,jorns}, of glucose metabolism
\cite{heimberg}, of changes in calcium concentration 
\cite{grapengiesser,herchuelz}, and of changes in electrical activity 
\cite{valdeolmillos}. Physiologists have suspected the importance of this 
variability \cite{pipeleers}. $\beta$ cell oscillations, or bursts, are more
complex than the simple oscillations of a van der Pol--FitzHugh--Nagumo type
relaxation oscillator, but the basic mechanism that diversity provokes the
emergence of oscillations remains the same, and provides a qualitative
explanation for the emergence of global oscillations in the $\beta$ cells of
the mammalian pancreas. Heterogeneity is the norm in biological excitable
media, so there may well be other instances awaiting discovery of this
mechanism in operation.

\section*{Acknowledgments}

It is a pleasure to acknowledge fruitful conversations I have held with 
Etelvina Andreu, Ehud Meron, Oreste Piro, Raquel Pomares, and Juanvi
S\'anchez-Andr\'es. I should like to thank Miguel Hern\'andez University 
of Elche, Spain, for the unforgettable experience of my time there in 1998, 
during which I carried out this research. I acknowledge the financial support 
of the Spanish Consejo Superior de Investigaciones Cient\'{\i}ficas.


\end{twocolumns}

\begin{thebibliography}{10}

\bibitem{meron}
E. Meron, Phys. Rep. {\bf 218},  1  (1992).

\bibitem{zaikin}
A.~N. Zaikin and A.~M. Zhabotinsky, Nature {\bf 225},  535  (1970).

\bibitem{truscott}
J.~E. Truscott and J. Brindley, Bull. Math. Biol. {\bf 56},  981  (1994).

\bibitem{glass}
{\em Theory of Heart: Biomechanics, Biophysics, and Nonlinear Dynamics of
  Cardiac Function}, edited by L. Glass, P. Hunter, and A. McCulloch (Springer,
   1991).

\bibitem{panfilov}
{\em Computational Biology of the Heart}, edited by A.~V. Panfilov and A.~V.
  Holden (Wiley, 1997).

\bibitem{meissner}
H.~P. Meissner, Nature {\bf 262},  502  (1976).

\bibitem{dean}
P.~M. Dean and E.~K. Matthews, Nature {\bf 219},  389  (1968).

\bibitem{perez}
M. P\'erez-Armend\'ariz, C. Roy, D.~C. Spray, and M.~V.~L. Bennett, Biophys. J.
  {\bf 59},  76  (1991).

\bibitem{smolen}
P. Smolen, J. Rinzel, and A. Sherman, Biophys. J. {\bf 64},  1668  (1993).

\bibitem{sherman}
A. Sherman, Bull. Math. Biol. {\bf 56},  811  (1994).

\bibitem{andreu}
E. Andreu, B. Soria, and J.~V. S{\'a}nchez-Andr{\'e}s, preprint  (1998).

\bibitem{vdp}
B. van~der Pol and J. van~der Mark, Phil. Mag. (7) {\bf 6},  763  (1928).

\bibitem{fitz1}
R.~A. FitzHugh, J. Gen. Physiol. {\bf 43},  867  (1960).

\bibitem{fitz2}
R.~A. FitzHugh, Biophys. J. {\bf 1},  445  (1961).

\bibitem{nagumo}
J.~S. Nagumo, S. Arimoto, and S. Yoshizawa, Proc. IREE Aust. {\bf 50},  2061
  (1962).

\bibitem{winfree}
A.~T. Winfree, Chaos {\bf 1},  303  (1991), {\bf 2}, 273 (1992).

\bibitem{quakeletter}
J.~H.~E. Cartwright, E. Hern\'andez-Garc\'{\i}a, and O. Piro, Phys. Rev. Lett.
  {\bf 79},  527  (1997).

\bibitem{quakebc}
J.~H.~E. Cartwright, V.~M. Egu\'{\i}luz, E. Hern\'andez-Garc\'{\i}a, and O.
  Piro, Int. J. Bifurcation \& Chaos {\bf 9},  2197  (1999).

\bibitem{rorsman}
P. Rorsman and G. Trube, J. Physiol. {\bf 374},  531  (1986).

\bibitem{kramer}
L. Kramer, F. Hynne, P. {Graae S{\o}rensen}, and D. Walgraef, Chaos {\bf 4},
  443  (1994).

\bibitem{huygens}
C. Huygens, J. des Scavans, No. 11 (16 March, 1665).

\bibitem{huygens1}
C. Huygens, J. des Scavans, No. 12 (23 March, 1665).

\bibitem{huygens2}
C. Huygens,  in {\em {\OE}uvres Compl{\`{e}}tes de Christiaan Huygens}
  (Societ{\'{e}} Hollandaise des Sciences, {1888--1950}), Vol.~17, p.\
  185.

\bibitem{winfree2}
A.~T. Winfree, J. Theor. Biol. {\bf 16},  15  (1967).

\bibitem{kuramoto}
Y. Kuramoto,  in {\em Proceedings of the International Symposium on
  Mathematical Problems in Theoretical Physics}, {\em Lecture Notes in
  Physics}, edited by H. Araki (Springer, 1975), pp.\ 420--422.

\bibitem{sakaguchi1}
H. Sakaguchi and Y. Kuramoto, Prog. Theor. Phys. {\bf 76},  576  (1986).

\bibitem{sakaguchi2}
H. Sakaguchi, S. Shinomoto, and Y. Kuramoto, Prog. Theor. Phys. {\bf 77},  1105
   (1987).

\bibitem{wiesenfeld}
K. Wiesenfeld, P. Colet, and S.~H. Strogatz, Phys. Rev. Lett. {\bf 76},  404
  (1996).

\bibitem{bar-eli}
K. Bar-Eli, Physica D {\bf 14},  242  (1985).

\bibitem{ermentrout}
G.~B. Ermentrout, Physica D {\bf 41},  219  (1990).

\bibitem{mirollo}
R.~E. Mirollo and S.~H. Strogatz, J. Stat. Phys. {\bf 60},  245  (1990).

\bibitem{braiman1}
Y. Braiman, W.~L. Ditto, K. Wiesenfeld, and M.~L. Spano, Phys. Lett. A {\bf
  206},  54  (1995).

\bibitem{braiman2}
Y. Braiman, J.~F. Lindner, and W.~L. Ditto, Nature {\bf 378},  465  (1995).

\bibitem{fgpv}
M. Feingold, D.~L. Gonz\'alez, O. Piro, and H. Viturro, Phys. Rev. A {\bf 37},
  4060  (1988).

\bibitem{wiesenfeld2}
K. Wiesenfeld {\it et~al.}, Phys. Rev. Lett. {\bf 72},  2125  (1994).

\bibitem{collins}
J.~J. Collins, C.~C. Chow, and T.~T. Imhoff, Phys. Rev. E {\bf 52},  3321
  (1995).

\bibitem{pikovsky}
A.~S. Pikovsky and J. Kurths, Phys. Rev. Lett. {\bf 78},  775  (1997).

\bibitem{salomon}
D. Salomon and P. Meda, Exp. Cell Res. {\bf 162},  507  (1986).

\bibitem{stefan}
Y. Stefan, P. Meda, M. Neufeld, and L. Orci, J. Clin. Invest. {\bf 80},  175
  (1987).

\bibitem{schuit}
F. Schuit, P. {In't Veld}, and D. Pipeleers, Proc. Nat. Acad. Sci. {\bf 85},
  3865  (1988).

\bibitem{hiriart}
M. Hiriart and M.~C. Ram\'{\i}rez-Medeles, Endrocrinology {\bf 128},  3193
  (1991).

\bibitem{jorns}
A. J{\"o}rns, Virchows Archiv {\bf 425},  305  (1994).

\bibitem{heimberg}
H. Heimberg {\it et~al.}, EMBO J. {\bf 12},  2873  (1993).

\bibitem{grapengiesser}
E. Grapengiesser, E. Gylfe, and B. Hellman, Arch. Biochem. Biophys. {\bf 268},
  404  (1989).

\bibitem{herchuelz}
A. Herchuelz, R. Pochet, C. Pastiels, and A. van Praet, Cell Calcium {\bf 12},
  577  (1991).

\bibitem{valdeolmillos}
M. Valdeomillos, A. Nadal, D. Contreras, and B. Soria, J. Physiol {\bf 455},
  173  (1992).

\bibitem{pipeleers}
D.~G. Pipeleers, Diabetes {\bf 392},  777  (1992).

\end{thebibliography}


\begin{references}
\bibitem[*]{jemail}
julyan@galiota.uib.es, WWW http://formentor.uib.es/$\sim$julyan.
\vspace{-0.4cm}
\def\references{}
\bibliographystyle{prsty}
\bibliography{database}
\end{references}
\end{document}